\documentclass[fleqn,usenatbib]{mnras}
\usepackage[T1]{fontenc}
\usepackage{ae,aecompl}
\usepackage{multicol}
\usepackage{longtable}
\usepackage{supertabular}
\usepackage{epsfig}
\usepackage{lineno}
\usepackage{lscape}
\usepackage{rotating}
\usepackage{natbib}
\usepackage{graphics}
\usepackage{gensymb}
\usepackage{graphicx}    
\usepackage{amssymb}    
\usepackage{float}
\hypersetup{draft}

\title[The distance to MAXI J1535--571]{An H\,{\sc i} absorption distance to the black hole candidate X-ray binary MAXI J1535--571}

\author[Chauhan et al.]{J. Chauhan$^{1}$\thanks{E-mail: j.chauhan@student.curtin.edu.au (JC)}, J.~C.~A. Miller-Jones$^{1}$, G.~E. Anderson$^{1}$, W. Raja$^{2}$, A. Bahramian$^{1}$,
\newauthor A. Hotan$^{2}$,  B. Indermuehle$^{2}$,  M. Whiting$^{2}$, J. R. Allison$^{3}$, C. Anderson$^{2}$, J. Bunton$^{2}$,
\newauthor B. Koribalski$^{2}$, E. Mahony$^{2}$\\
$^{1}$International Centre for Radio Astronomy Research -- Curtin University, GPO Box U1987, Perth, WA 6845, Australia\\
$^{2}$CSIRO Astronomy and Space Science, Australia Telescope National Facility, PO Box 76, Epping NSW 1710, Australia\\
$^{3}$Sub-Dept. of Astrophysics, Department of Physics, University of Oxford, Denys Wilkinson Building, Keble Rd., Oxford, OX1 3RH, UK\\
}
\date{Accepted XXX. Received YYY; in original form ZZZ}

\pubyear{2019}

\begin{document}
\label{firstpage}
\pagerange{\pageref{firstpage}--\pageref{lastpage}}
\maketitle

\begin{abstract}
With the Australian Square Kilometre Array Pathfinder (ASKAP) we monitored the black hole candidate X-ray binary MAXI J1535--571 over seven epochs from 21 September to 2 October 2017. Using ASKAP observations, we studied the H\,{\sc i} absorption spectrum from gas clouds along the line-of-sight and thereby constrained the distance to the source. The maximum negative radial velocities measured from the H\,{\sc i} absorption spectra for MAXI J1535--571 and an extragalactic source in the same field of view are $-69\pm4$\,km\,s$^{-1}$ and $-89\pm4$\,km\,s$^{-1}$, respectively. This rules out the far kinematic distance ($9.3^{+0.5}_{-0.6}$\,kpc), giving a most likely distance of $4.1^{+0.6}_{-0.5}$\,kpc, with a strong upper limit of the tangent point at $6.7^{+0.1}_{-0.2}$\,kpc. At our preferred distance, the peak unabsorbed luminosity of MAXI J1535--571  was $>78$ per cent of the Eddington luminosity, and shows that the soft-to-hard spectral state transition occurred at the very low luminosity of 1.2 -- 3.4 $\times$ 10$^{-5}$ times the Eddington luminosity. Finally, this study highlights the capabilities of new wide-field radio telescopes to probe Galactic transient outbursts, by allowing us to observe both a target source and a background comparison source in a single telescope pointing.
\end{abstract}

\begin{keywords}
black hole physics -- ISM: jets and outflows -- radio continuum: transients -- X-rays: binaries -- X-rays: individual: MAXI J1535--571
\vspace{-3.0em}
\end{keywords}

\renewcommand{\thefootnote}{\arabic{footnote}}
\section{Introduction}
Stellar-mass black holes in binary systems play an important role in understanding the universal phenomena of accretion and ejection \citep{Fender2004}, as they evolve on humanly observable time-scales. However, black hole X-ray binaries (XRBs) typically have poorly constrained distances, with uncertainties often more than 50\% \citep{Jonker2004}. The distance of a stellar object is of great importance as it enables us to determine the physical parameters from observables, such as luminosity from flux, physical separation from angular separation, and velocity from proper motion. 

Parallax is the most fundamental and model-independent method for estimating stellar distances. The main caveat of this method is that for typical XRB distances, the measured parallax is extremely small (1 milliarcsec for 1 kpc). Therefore, instruments capable of probing sub-milliarcsecond scales are required. Such studies are possible only with very long baseline interferometry (VLBI) and {\it Gaia} \citep{Gaia2018}, and only for a handful of XRBs \citep[e.g.][]{Miller-Jones2009,Reid2011, Gandhi2019}.

High-precision trigonometric parallax measurements are not possible for most black hole XRBs. In such cases, the distance is commonly estimated by comparing the inferred absolute magnitude of the donor star (as determined from its spectral type) with the apparent magnitude, after considering the maximum contribution from the accretion disk \citep{Jonker2004}.  The main limitations of this technique are the uncertainties associated with the disk contribution and the extinction along the line of sight.

If the companion star is too faint (e.g. if the system is highly extincted) then other methods must be used, one of which uses the Doppler shifted 21-cm absorption line of neutral hydrogen \citep[H\,{\sc i;} e.g.][]{Koribalski1995, Dhawan2000, Lockman2007}. Intervening interstellar H\,{\sc i} clouds are presumed to follow the rotation of the Milky Way and their absorption features allow us to constrain the kinematic distance to the source \citep{Gathier1986, Kuchar1990}. If the source is beyond the solar orbit, this method provides an accurate kinematic distance estimation. However, the method suffers from ambiguities if the source lies within the solar orbit \citep{Wenger2018}.

On 2 September 2017, an uncatalogued XRB was co-discovered by the Monitor of All-sky X-ray Image\footnote{\url{http://maxi.riken.jp/top/index.html}} \citep[{\it MAXI};][]{Matsuoka2009} and the Neil Gehrels Swift Observatory \citep[{\it Swift};][]{Gehrels2004}, and designated MAXI J1535--571 \citep{Negoro2017a, Kennea2017a}. In J2000.0 coordinates, the source is located at RA = 15:35:19.714$\pm0.007$, DEC = --57:13:47.583$\pm0.024$ \citep[Galactic  coordinates l\,=\,323.72407\degree$, b\,=\,$-$1.12887\degree$;][]{Russell2017a}. Multi-wavelength studies of the system strongly suggest that the accreting compact object is a stellar-mass black hole \citep{Dincer2017, Kennea2017b, Negoro2017b, Russell2017a, Russell2017b}. However, the physical parameters, including the black hole mass and distance, and the peak luminosity of the outburst are still uncertain.

In this study, we aim to constrain the distance to MAXI J1535--571 using H\,{\sc i} absorption spectra from Australian Square Kilometre Array Pathfinder (ASKAP) early science observations. ASKAP is a synthesis radio telescope located at the Murchison Radio-astronomy Observatory (MRO) in Western Australia, which has already established its ability to detect H\,{\sc i} absorption from extragalactic sources \citep{Allison2015, Allison2017}. The telescope operates in the frequency range 0.7--1.8 GHz providing an angular resolution of 10 arcsec and a field of view (FOV) of 30 deg$^{2}$ at 1.4 GHz \citep{Hotan2014}. This allows us to simultaneously measure the H\,{\sc i} absorption towards both MAXI J1535--571 and an extragalactic calibrator source, allowing us to distinguish between near and far kinematic distances.

\vspace{-2.0em}
\section{Observations and Data reduction}
During the 2017 outburst of MAXI J1535--571, ASKAP observed the source over seven epochs from 21 September to 2 October 2017. At the time of these early science observations, ASKAP was operated in a sub-array of twelve dishes, which provides an angular resolution of $\sim30$ arcsec. All observations were conducted at a central frequency of 1.34 GHz with a processed bandwidth of 192 MHz. The total bandwidth was further divided into 10368 fine channels, each with a frequency resolution of 18.519 kHz, with visibilities recorded every 10 s. Further details of the observations can be seen in Table \ref{tab:tab1}.

\begin{table*}
 \centering
 \caption{Details of our ASKAP observations of MAXI J1535--571.} 
\begin{center}
\scalebox{0.75}{%
\begin{tabular}{ |l|c|c|c|c|c|c|c|c|c| }
 \hline
\hline
Observation & Observation & Observation & MJD$^{\mathrm{b}}$ & Exposure & Flux Density$^{\mathrm{a}}$ & rms noise\\
ID  & Start date & Start time & & time & $S_{0}$ & (mJy\,beam$^{-1}$) \\
 & (DD-MM-YYYY) & (hh:mm:ss) & & (hh:mm:ss) & (mJy) & \\
\hline
4340 & 21-09-2017 & 03:06:32 & 58017.17 & 02:02:26 & $579.6\pm2.1$ &  0.74\\
4364 & 22-09-2017 & 01:30:02 & 58018.12 & 02:51:50 & $156.1\pm1.9$ &  0.72\\
4367 & 22-09-2017 & 10:35:41 & 58018.50 & 03:00:56 & $306.1\pm1.3$ &  0.54\\
4374 & 23-09-2017 & 13:23:39 & 58019.58 & 01:00:52 & $478.2\pm2.4$ &  0.91\\
4410 & 30-09-2017 & 03:29:59 & 58026.23 & 04:00:56 & $\phantom{1}39.8\pm0.8$ &  0.53\\
4414 & 01-10-2017 & 03:35:00 & 58027.23 & 04:01:01 & $\phantom{1}26.3\pm0.8$ &  0.54\\
4418 & 02-10-2017 & 03:39:59 & 58028.24 & 04:01:22 & $\phantom{1}21.4\pm0.7$ &  0.52\\
\hline
\hline
\end{tabular}}\\
\end{center}
\begin{flushleft}
$^{\mathrm{a}}$ $1\sigma$ errors are quoted, calculated by adding in quadrature the $1\sigma$ error on the Gaussian fit and the $1\sigma$ rms noise in the image.\\
$^{\mathrm{b}}$ Mid point of our observations.\\
\end{flushleft}
\vspace{-3.5em}
\label{tab:tab1}
\end{table*}

We reduced our data using the standard ASKAP data analysis software,  {\tt ASKAPsoft}\footnote{\url{http://www.atnf.csiro.au/computing/software/askapsoft/sdp/docs/current/index.html}} (pipeline version 0.24.1). For continuum imaging, we performed dynamic flagging of faulty antennas, faulty channels, and baselines affected by radio frequency interference (RFI) during bandpass calibration. Bandpass and flux calibration were performed using the calibrator source PKS B1934--638, and the bandpass solution was interpolated across frequencies where data were flagged. The {\tt IMFIT} task in the Common Astronomy Software Application \citep[{\sc CASA} v5.1.2-4:][]{McMullin2007} package was used for measuring the flux densities and $1\sigma$ uncertainties of MAXI J1535--571 from the continuum images, which are listed in Table \ref{tab:tab1}.

With a flux density $>150$ mJy during the first four epochs, there was enough signal-to-noise to detect H\,{\sc i} absorption towards MAXI J1535--571. We used the {\tt SIMAGER} task to generate spectral-line cubes, and performed only stokes-V flagging to avoid removing the true absorption line in the H\,{\sc i} region. We applied the bandpass solutions to the raw spectral data and extracted a sub-spectral cube with 1000 channels centered at 1.42 GHz and a velocity resolution of $\sim4$ km/s. Note that our conclusions should not be affected by the Galactic H\,{\sc i} absorption towards PKS B1934--638, which is very narrow (FWHM $<$ 6 km/s; approximately one channel wide) and therefore only affects the H\,{\sc i} absorption line at local standard of rest (LSR) velocity\,=\,0 km s$^{-1}$ \citep{Denes2018}. We extracted the H\,{\sc i} spectra corresponding to MAXI J1535--571 and a nearby extragalactic comparison source PMN J1533--5642 \citep[= MGPS J153358--564218;][]{Griffith1993, Murphy2007}, located at RA\,=\,15:33:57.439, DEC\,=\,$-$56:42:17.183 (uncertainty of 10 arcsec), and analysed the H\,{\sc i} absorption complexes with significance exceeding $3\sigma$.

\vspace{-2.0em}
\section{Results}
\begin{figure*}
    \includegraphics[width=\columnwidth]{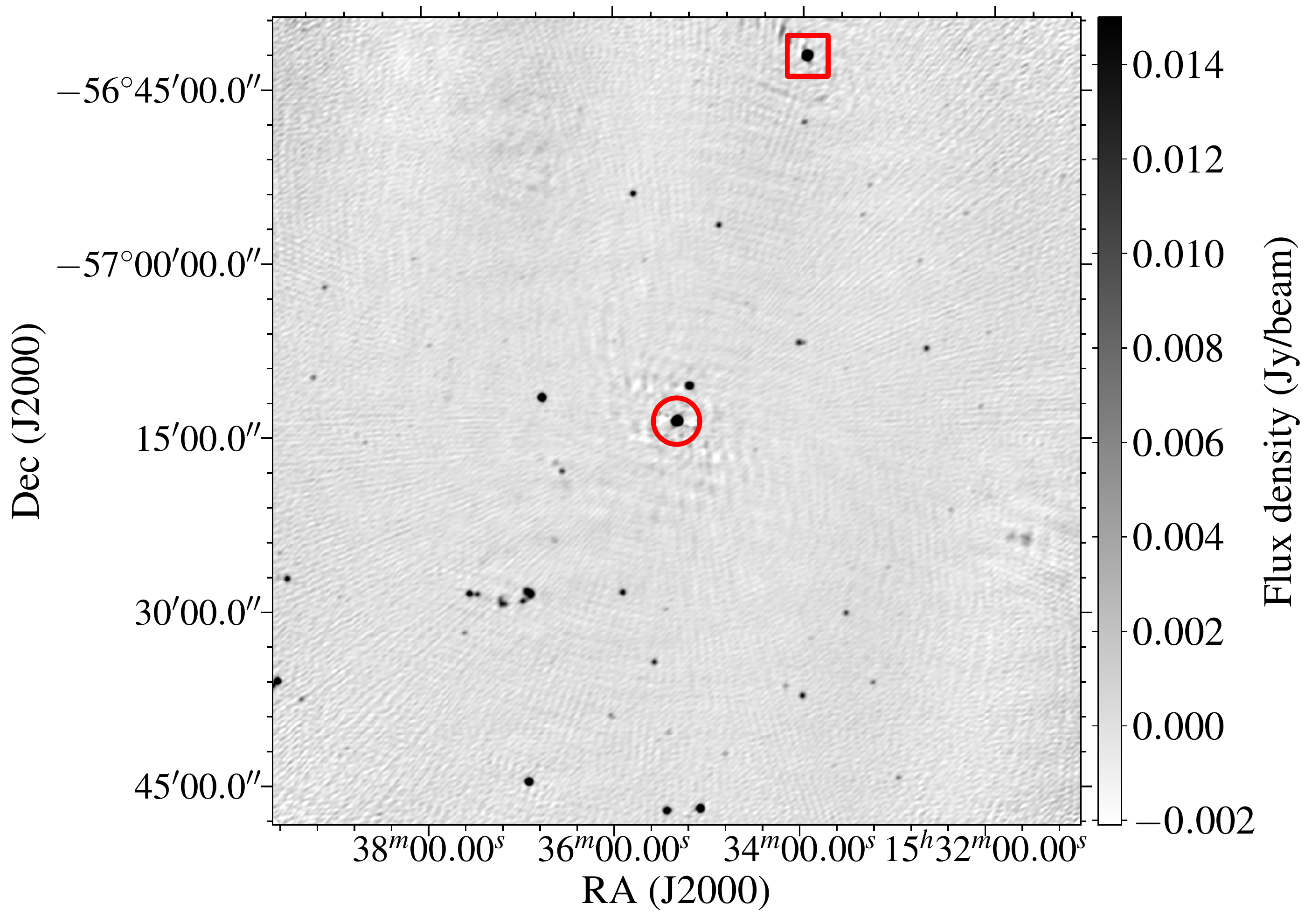}
    \includegraphics[width=0.75\columnwidth]{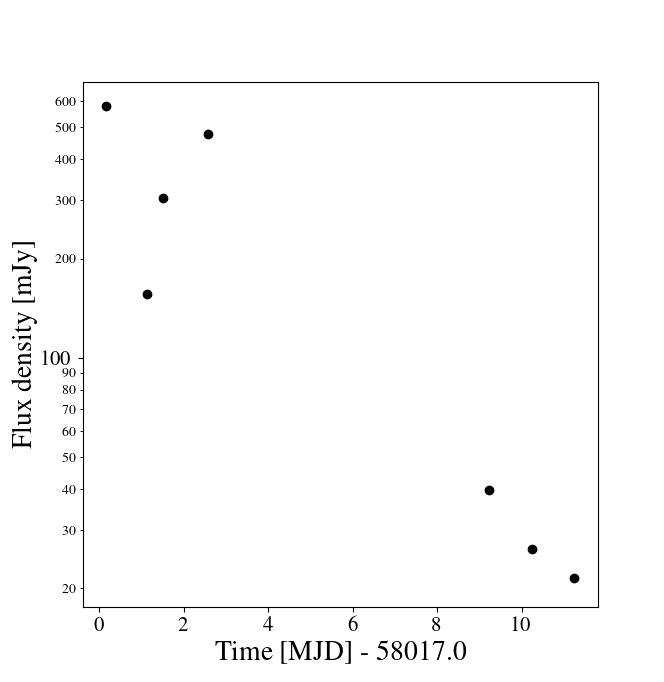}\vspace{-1.0em}
    \caption{{\bf Left} panel: 1.34 GHz ASKAP continuum image of our 21 September 2017 observation centered at RA = 15:35:19.71, DEC = --57:13:47.58 with a size of 1.15$\degree$ $\times$ 1.15$\degree$. The positions of MAXI J1535--571 and the comparison extragalactic source MGPS J153358--564218 are indicated by the red circle and square, respectively. {\bf Right} panel: ASKAP light curve of MAXI J1535--571 at 1.34 GHz. The peak flux density of MAXI J1535--571 exceeded 500 mJy.} 
    \vspace{-2.0em}
    \label{fig:fig1}
\end{figure*}

The continuum image of the region around MAXI J1535--571 for the 21 September 2017 observation is shown in Figure \ref{fig:fig1} (left panel), where the source is detected at high significance ($>200\sigma$). Residual calibration, deconvolution and amplitude errors remain in our images due to these data being obtained during ASKAP's early science phase with 12 of 36 active antennas, causing reduced UV coverage, the use of the rapidly evolving {\tt ASKAPsoft} software, and the lack of complete instrumental characterisation that will only improve over time. Figure \ref{fig:fig1} (left panel) also shows the extragalactic comparison source MGPS J153358--564218 with a measured flux density of $340\pm2$ mJy on 21 September 2017, which remained stable to better than 10\% between all the epochs.

In the right panel of Figure \ref{fig:fig1}, we present the 1.34 GHz light curve of MAXI J1535--571 from 21 September to 2 October 2017. During this period we observed two distinct peaks on 21 and 23 September, reaching flux densities $>450$ mJy, likely corresponding to separate flaring events. Subsequently, the radio flux density of MAXI J1535--571 gradually decreased, reaching a level of $\sim20$ mJy by 2 October. 

\begin{figure}
\vspace{-1.5em}
    \includegraphics[width=0.9\columnwidth]{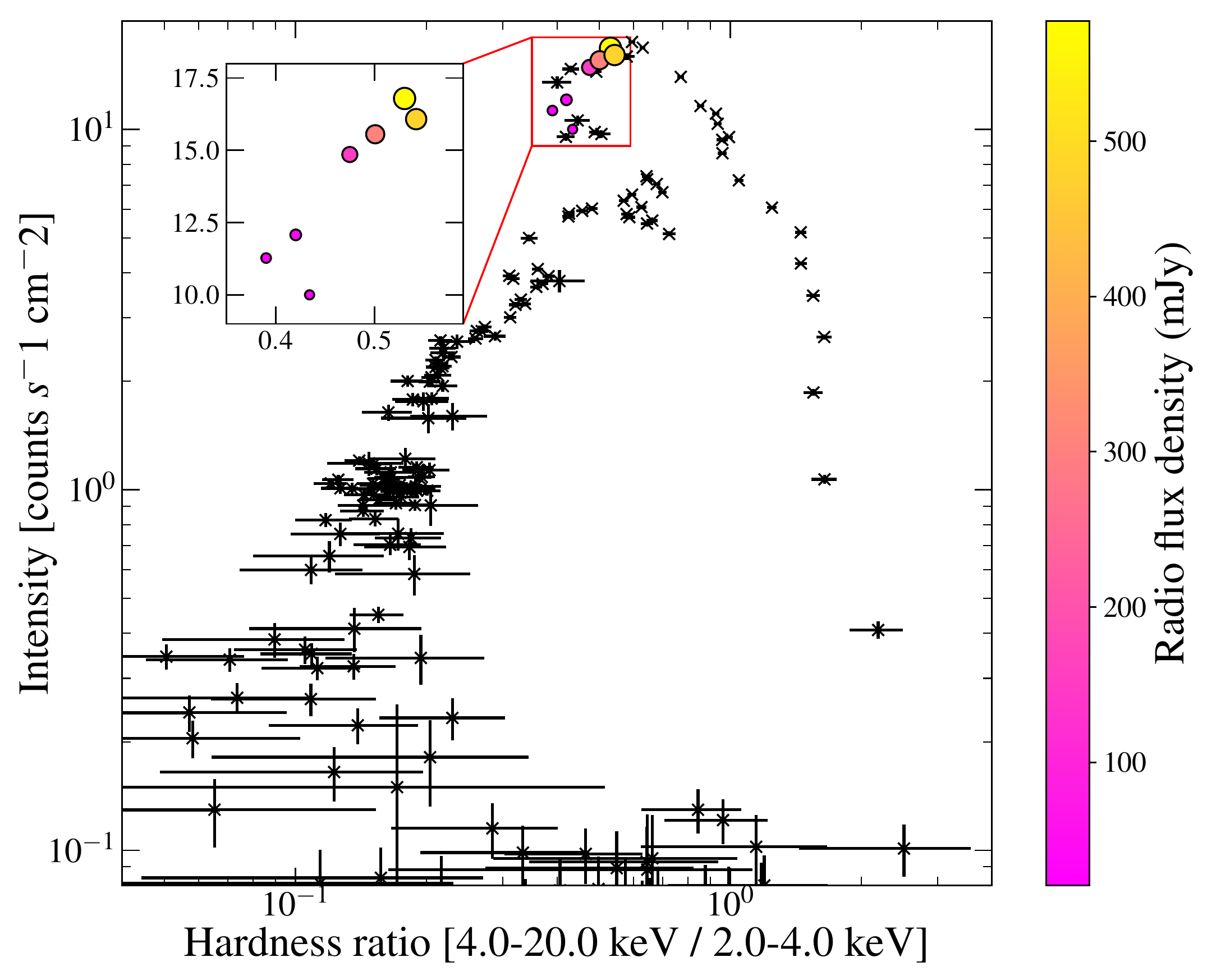}\vspace{-1.0em}
    \caption{Hardness intensity diagram (HID) for the 2017 outburst of MAXI J1535--571. The HID is made from public {\it MAXI} data, where intensity is the count rate in the energy range 2.0--20.0 keV, and the hardness ratio is the ratio of count rates in the 4.0--20.0 keV and 2.0--4.0 keV energy bands \citep{Tao2018}. Those X-ray data-points with simultaneous radio detections are highlighted with circles, whose size and color correspond to the measured radio flux density (zoomed in the inset of the plot). The radio flare occurred  during the hard-to-soft X-ray spectral state transition.}
    \vspace{-1.0em}
    \label{fig:fig2}
\end{figure}

We downloaded the on-demand X-ray light curves from {\it MAXI} in the energy range 2.0--20.0 keV to create a hardness-intensity diagram (HID) for MAXI J1535--571, presented in Figure \ref{fig:fig2}. From the HID, all our radio observations were taken in the soft-intermediate spectral state \citep{Tao2018}. This confirms that MAXI J1535--571 underwent radio flaring behaviour during the hard-to-soft X-ray spectral state transition, as is typical for XRBs \citep{Fender2004}.
\vspace{-2.0em}
\subsection{H\,{\sc i} absorption spectra \label{sec:HIabsspec}}
\begin{figure}
\vspace{-1.8em}
   \includegraphics[width=\columnwidth]{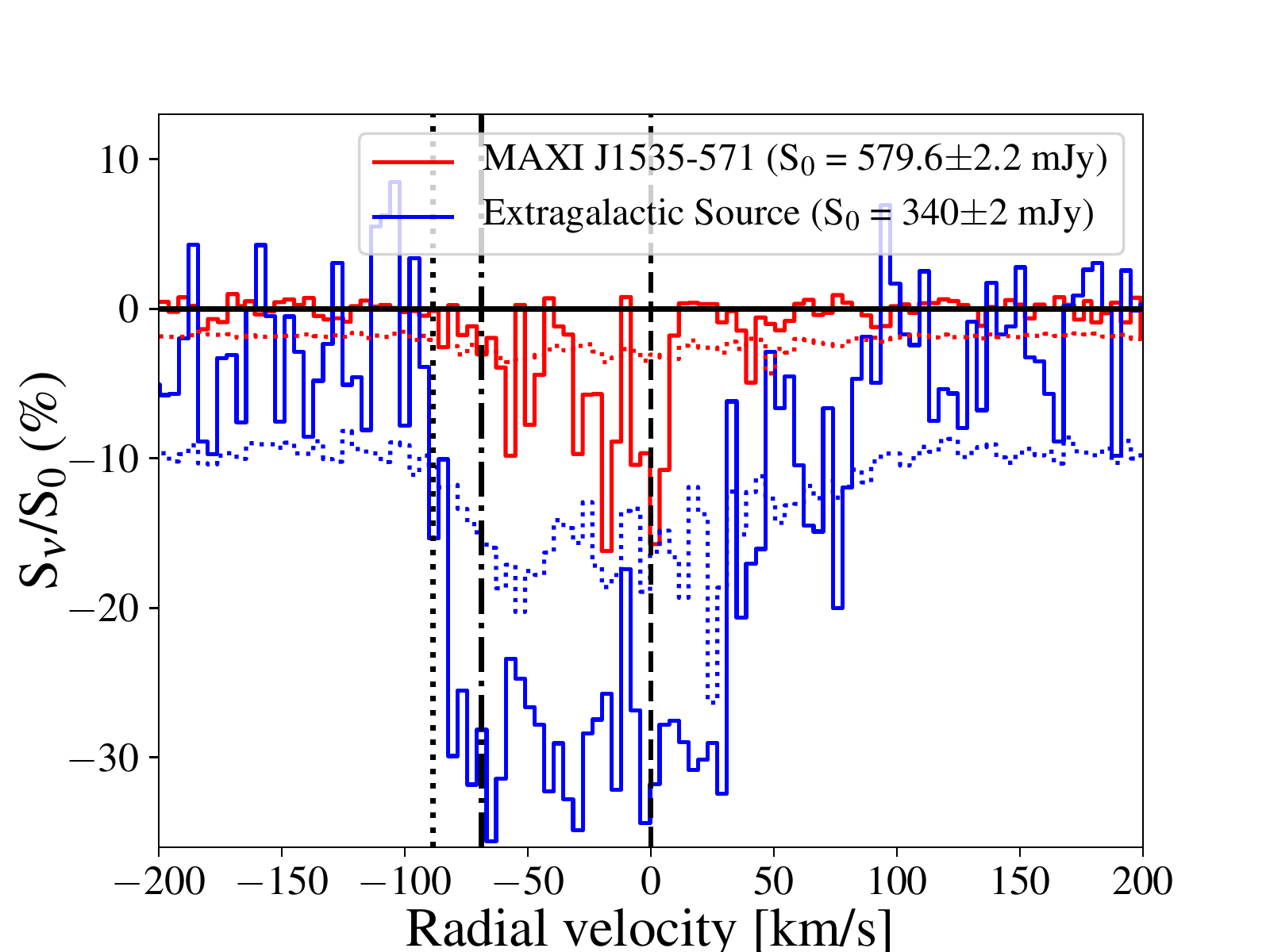}\vspace{-2.0em}
    \caption{The H\,{\sc i} absorption spectrum observed from MAXI J1535--571 on 21 September 2017, when the source was brightest ($S_{0}=579.6\pm2.1$\,mJy). Spectra of MAXI J1535--571 and the extragalactic source MGPS J153358--564218 (in the local standard of rest frame) are displayed in red and blue, respectively. $S_{\nu}$ is measured from the data cube and $S_{0}$ is determined from the continuum image. The solid black horizontal line shows the zero flux level. The red and blue dotted lines represent the $3\sigma$ rms noise levels (calculated directly from individual channel image in the spectral cube) for MAXI J1535--571 and the extragalactic source, respectively. The latter is higher due to being located further out in the primary beam. The dashed vertical line shows the rest velocity. The maximum negative velocity of significant ($>3\sigma$) absorption in the MAXI J1535--571 spectrum (dash-dotted vertical line at $-69$\,km\,s$^{-1}$) places a lower limit on the source distance of $4.1^{+0.6}_{-0.5}$\,kpc. The maximum negative velocity derived from the extragalactic source spectrum (dotted vertical line at $-89$\,km\,s$^{-1}$) constrains the upper limit on the distance of MAXI J1535--571 to be that of the tangent point at $6.7^{+0.1}_{-0.2}$}\,kpc.
    \vspace{-1.7em}
    \label{fig:fig3}
\end{figure}

In Figure \ref{fig:fig3}, we show the H\,{\sc i} spectrum for MAXI J1535--571 for the first epoch (21 September 2017). In the same plot, we also display the spectrum for the extragalactic source MGPS J153358--564218, and the $3\sigma$ rms noise levels for the spectra of both sources. For both sources, we measured the rms noise per channel from the spectral cube, taking the median noise value in each channel from a number of regions close to the source to mitigate against any faint Galactic H{\sc i} emission in any given region. The rms noise level for both sources is shown as the dotted line (with their respective red and blue colours) in Figure \ref{fig:fig3}, where $S_{\nu}$ is the flux density measured from the data cube and $S_{0}$ is that measured from the continuum image via the {\tt IMFIT} task. In the case of MAXI J1535--571, we observed a significant ($>3\sigma$) H\,{\sc i} absorption complex with a maximum negative velocity of $-69\pm4$ km s$^{-1}$. In the second epoch, the H\,{\sc i} absorption lines are weaker due to the lower flux density. Therefore, we have not used the second epoch because the sensitivity was not high enough to reliably determine the maximum velocity out to which absorption could be observed. The third and fourth epochs are excluded from this analysis due to the presence of RFI in the H\,{\sc i} spectrum.

\begin{figure}
\vspace{-1.8em}
    \includegraphics[width=0.9\columnwidth]{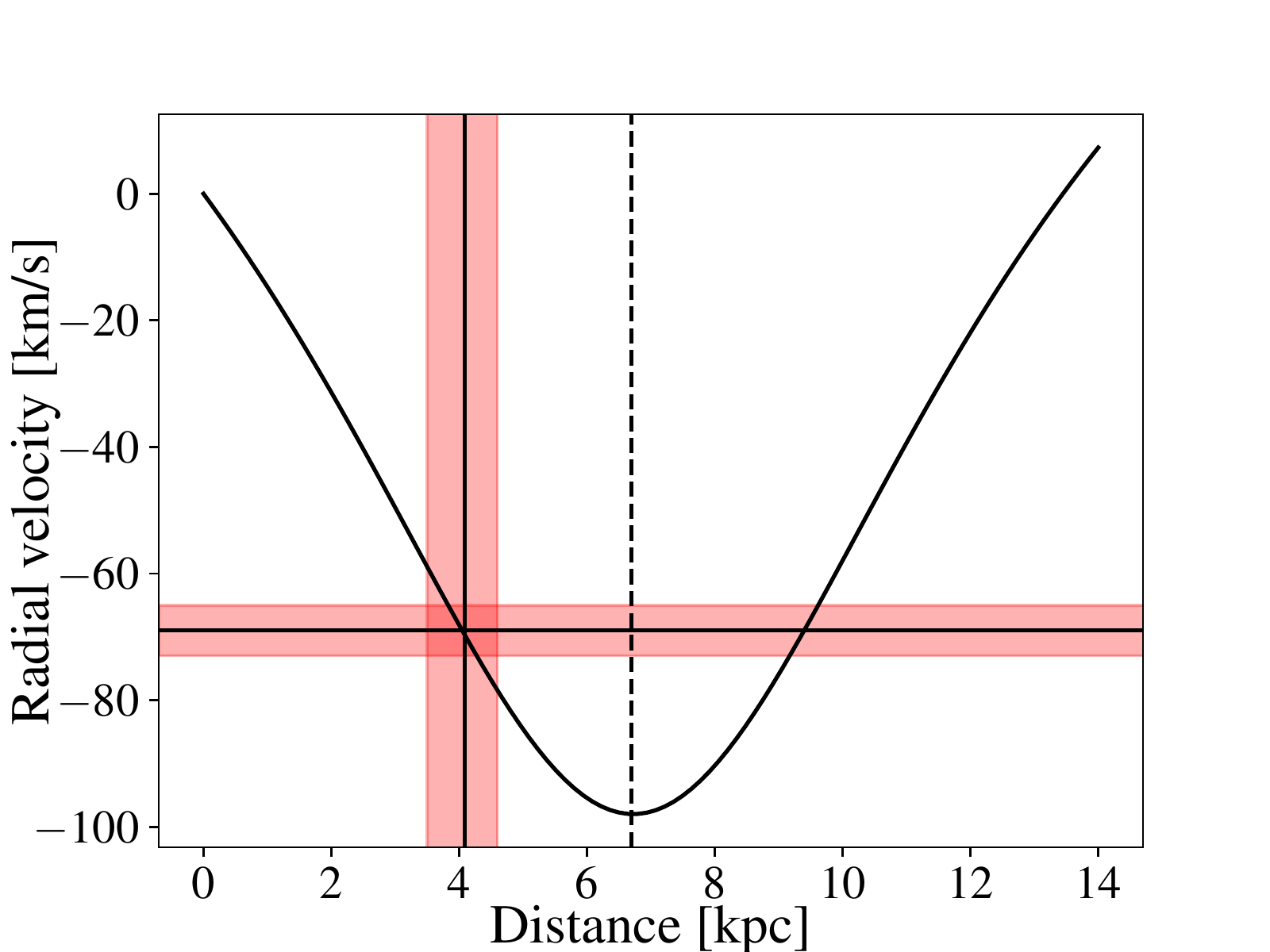}\vspace{-1.0em}
    \caption {The variation of the predicted velocity (for circular rotation) of the local standard of rest ($V_{\rm LSR}$) as a function of distance from the Sun ($d$), in the direction of MAXI J1535--571. The horizontal line shows the velocity of $-69\pm4$ km s$^{-1}$, measured from the H\,{\sc i} absorption complex of the first epoch. The vertical line displays the most likely distance ($4.1^{+0.6}_{-0.5}$\,kpc) to MAXI J1535--571. The shaded region across both the lines highlight the respective error regions, as derived from the Monte Carlo approach of \citet{Wenger2018}. The dashed line displays the position of the tangent point, which is our robust distance upper limit of $6.7^{+0.1}_{-0.2}$\,kpc.}
    \vspace{-1.5em}
    \label{fig:fig5}
\end{figure}

For Galactic longitudes $270\degree\leq l \leq360\degree$, the expected radial velocities from Galactic rotation are negative. Figure \ref{fig:fig5} shows the calculated variation of velocity of the local standard of rest ($V_{\rm LSR}$) with distance from the Sun ($d$) in the direction of MAXI J1535--571. For the determination of $V_{\rm LSR}$, we presumed the Galactic rotation curve to be flat with a circular velocity ($V_{0}$) and a Galactic centre distance ($R_{0}$) of $240\pm8$\,km\,s$^{-1}$ and $8.34\pm0.16$\,kpc, respectively \citep{Reid2014}. $V_{\rm LSR}$ is symmetric about the tangent point distance ($R = R_{0}\cos{l}\equiv6.7^{+0.1}_{-0.2}$ kpc), and therefore provides two estimates of the distance, near and far, within the solar circle. Beyond $d=13.2$\,kpc, where $R>R_{0}$, $V_{\rm LSR}$ becomes positive. The kinematic distance to MAXI J1535--571 suffers ambiguities because the source is located within the solar orbit.  Furthermore, the gas clouds in the Milky Way may not all be following circular motions. Therefore, we have followed the prescription given by \citet{Wenger2018}, who used the rotation curve given by \citet{Reid2014} with a Monte Carlo approach\footnote{\url{http://www.treywenger.com/kd/index.php}} to derive a more robust kinematic distance and uncertainty. Using the observed maximum negative absorption velocity and the methodology of \citet{Wenger2018}, we constrain the near and far source distances to be $4.1^{+0.6}_{-0.5}$\,kpc and $9.3^{+0.5}_{-0.6}$\,kpc, respectively.

The H\,{\sc i} absorption spectrum of the extragalactic source MGPS J153358--564218 (Figure \ref{fig:fig3}) shows $>3\sigma$ significance absorption out to a maximum negative velocity of $-89\pm4$\,km\,s$^{-1}$, which is due to more distant gas clouds, and consistent (within uncertainties) with the tangent point velocity of $-92^{+8}_{-9}$\,km\,s$^{-1}$ (obtained following \citealt{Reid2014} and \citealt{Wenger2018}). Absorption at the tangent point velocity is not seen towards MAXI J1535--571, implying that MAXI J1535--571 should be located closer than the tangent point distance of $6.7^{+0.1}_{-0.2}$\,kpc (Figure \ref{fig:fig5}).
\vspace{-2.0em}
\section{Discussion}
As outlined in section \ref{sec:HIabsspec}, we place a lower limit of $4.1^{+0.6}_{-0.5}$\,kpc on the distance to MAXI J1535--571. The more negative radial velocities detected towards the extragalactic source must therefore be coming from larger distances, out to at least the tangent point at $6.7^{+0.1}_{-0.2}$ kpc. If we discount the unlikely scenario that all absorption at intermediate velocities comes from gas clouds more distant than the tangent point, we find a most probable distance of $4.1^{+0.6}_{-0.5}$\,kpc for MAXI J1535--571. In the following subsections we discuss the implications of these distance constraints.

\vspace{-1.0em}
\subsection{Spectral state transition luminosity}
According to \citet{Kalemci2013}, the soft-to-hard spectral state transition in black hole XRBs usually occurs between 0.3 and 3 per cent of the Eddington luminosity. A {\it Swift}/XRT observation on 29 April 2018 was identified as the point of the soft-to-hard spectral state transition \citep{Tao2018}, with an unabsorbed X-ray flux of $\sim6$ $\times$ 10$^{-12}$\,erg\,cm$^{-2}$\,s$^{-1}$. Using our distance estimates, and assuming a typical black hole mass of $8\pm1 M_{\odot}$ \citep{Kreidberg2012}, we calculate a transition luminosity of 1.2 -- 3.4 $\times$ 10$^{-5}$ times the Eddington luminosity, which is much lower than is typical for black hole XRBs \citep{Kalemci2013}.

Such a peculiar behavior has been seen in a few other sources, and there are a few possible explanations. \citet{Vahdat2019} hypothesised that a new episode of accretion during an outburst decay could launch a new soft state, with a lower transition luminosity than the normal 0.3 -- 3 \% of the Eddington luminosity. Alternatively, \citet{Petrucci2008} and \citet{Begelman2014} suggested that low viscosity or low magnetic fields in the accretion disk could also decrease the state transition luminosity.

\vspace{-1.0em}
\subsection{Peak X-ray luminosity}
\citet{Stiele2018} observed a peak absorbed X-ray flux of $\sim3.7$ $\times$ 10$^{-7}$\,erg\,cm$^{-2}$\,s$^{-1}$ from MAXI J1535--571 in the energy range 0.6--10 keV. Using our derived distance of $4.1^{+0.6}_{-0.5}$\,kpc, this corresponds to an absorbed luminosity of 7.5 -- 19.9 $\times$ 10$^{38}$\,erg\,s$^{-1}$. For a compact object of typical black hole mass $8\pm1 M_{\odot}$ \citep{Kreidberg2012}, this corresponds to $>78$ per cent of the Eddington luminosity. \citet{Tao2018} reported that over the peak of the outburst, the X-ray spectrum could be fit with a slim disk model, suggesting a luminosity close to the Eddington luminosity. Their conclusion is therefore consistent with our distance estimate.

\vspace{-1.0em}
\subsection{Line-of-sight hydrogen column density}
From our distance constraints and Galactic latitude ($b$), we infer MAXI J1535--571 to lie 0.08--0.13\,kpc below the Galactic plane, well within the scale height of the Milky Way disk \citep[$\sim0.3$\,kpc;][]{Mandel2016}. Using X-ray spectroscopy, the absorption column density ($n_{\rm H}$) towards MAXI J1535--571 was found to be in the range 2.8 -- 5.6 $\times 10^{22}$\,cm$^{-2}$ \citep{Miller2018, Stiele2018, Tao2018}, which is consistent with the predicted Galactic absorption value of $2.0\pm1.0 \times 10^{22}$\,cm$^{-2}$ in the direction of MAXI J1535--571, calculated from the extinction maps and estimated correlation between $n_{\rm H}$ and $A_{\rm V}$ \citep{Schlafly2011, Bahramian2015, Foight2016}.

\vspace{-2.5em}
\section{Summary}
In this study, we measured a kinematic distance to the black hole candidate XRB MAXI J1535--571 using an H\,{\sc i} absorption spectrum obtained from ASKAP early science observations. A comparison extragalactic source in the same field shows H\,{\sc i} absorption out to $-89\pm4$\,km\,s$^{-1}$ (consistent with coming from the tangent point at $6.7^{+0.1}_{-0.2}$\,kpc), whereas MAXI J1535--571 shows no significant H\,{\sc i} absorption beyond $-69\pm4$\,km\,s$^{-1}$. Therefore, the most probable distance to the source is $4.1^{+0.6}_{-0.5}$\,kpc, with a robust upper limit of $6.7^{+0.1}_{-0.2}$\,kpc.

Our distance constraints have confirmed that MAXI J1535--571 was accreting close to the Eddington rate at the peak of the outburst, in agreement with X-ray spectral fitting results by \citet{Tao2018}. We also demonstrate that the soft-to-hard X-ray spectral state transition in MAXI J1535--571 occurred at a very low fraction (1.2 -- 3.4 $\times$ 10$^{-5}$) of the Eddington luminosity, which could either be due to a second episode of accretion partway through the outburst decay, or to a low viscosity and magnetic field in the accretion disk.

This study also highlights the importance of modern radio telescopes such as ASKAP, which provide both high sensitivity and a wide field-of-view. In this project, the wide field-of-view of ASKAP afforded us the opportunity to study the H\,{\sc i} absorption spectrum of both MAXI J1535--571 and a comparison extragalactic source within the same observation. The high sensitivity and high spectral resolution of ASKAP allowed us to extract the H\,{\sc i} spectrum from just two hours of observation, placing stringent constraints on the distance to MAXI J1535--571.

\vspace{-2.5em}
\section*{Acknowledgements}
We thank the referee for their valuable comments. The Australian SKA Pathfinder is part of the Australia Telescope National Facility which is managed by CSIRO. Operation of ASKAP is funded by the Australian Government with support from the National Collaborative Research Infrastructure Strategy. ASKAP uses the resources of the Pawsey Supercomputing Centre. Establishment of ASKAP, the Murchison Radio-astronomy Observatory and the Pawsey Supercomputing Centre are initiatives of the Australian Government, with support from the Government of Western Australia and the Science and Industry Endowment Fund. We acknowledge the Wajarri Yamatji people as the traditional owners of the Observatory site. JCAM-J is the recipient of Australian Research Council Future Fellowship (project number FT140101082) and GEA is the recipient of an Australian Research Council Discovery Early Career Researcher Award (project number DE180100346) funded by the Australian Government.



\bsp    
\label{lastpage}

\vspace{-2.5em}
\begin{thebibliography}{99}
\bibitem[\protect\citeauthoryear{Allison et al.}{2015}]{Allison2015} Allison J.~R., et al., 2015, MNRAS, 453, 1249
\bibitem[\protect\citeauthoryear{Allison et al.}{2017}]{Allison2017} Allison J.~R., et al., 2017, MNRAS, 465, 4450
\bibitem[\protect\citeauthoryear{Bahramian et al.}{2015}]{Bahramian2015} Bahramian A., Heinke C.~O., Degenaar N., Chomiuk L., Wijnands R., Strader J., et al., 2015, MNRAS, 452, 3475
\bibitem[\protect\citeauthoryear{Begelman \& Armitage}{2014}]{Begelman2014} Begelman M.~C., Armitage P.~J., 2014, ApJ, 782, L18
\bibitem[\protect\citeauthoryear{D{\'e}nes et al.}{2018}]{Denes2018} D{\'e}nes H., McClure-Griffiths N.~M., Dickey J.~M., Dawson J.~R., Murray C.~E., 2018, MNRAS, 479, 1465
\bibitem[\protect\citeauthoryear{Dhawan, Goss, \& Rodr{\'{\i}}guez}{2000}]{Dhawan2000} Dhawan V., Goss W.~M., Rodr{\'{\i}}guez L.~F., 2000, ApJ, 540, 863
\bibitem[\protect\citeauthoryear{Dincer}{2017}]{Dincer2017} Dincer T., 2017, ATel1, 10716
\bibitem[\protect\citeauthoryear{Fender, Belloni, \& Gallo}{2004}]{Fender2004} Fender R.~P., Belloni T.~M., Gallo E., 2004, MNRAS, 355, 1105
\bibitem[\protect\citeauthoryear{Foight, G{\"u}ver, {\"O}zel \& Slane}{2016}]{Foight2016} Foight D.~R., G{\"u}ver T., {\"O}zel F., Slane P.~O., 2016, ApJ, 826, 66
\bibitem[\protect\citeauthoryear{Gaia Collaboration et al.}{2018}]{Gaia2018} Gaia Collaboration, et al., 2018, A\&A, 616, A1
\bibitem[\protect\citeauthoryear{Gandhi et al.}{2019}]{Gandhi2019} Gandhi P., Rao A., Johnson M.~A.~C., Paice J.~A., Maccarone T.~J., 2019, MNRAS, 485, 2642
\bibitem[\protect\citeauthoryear{Gathier, Pottasch, \& Goss}{1986}]{Gathier1986} Gathier R., Pottasch S.~R., Goss W.~M., 1986, A\&A, 157, 191
\bibitem[\protect\citeauthoryear{Gehrels, et al.}{2004}]{Gehrels2004} Gehrels N., et al., 2004, ApJ, 611, 1005
\bibitem[\protect\citeauthoryear{Griffith \& Wright}{1993}]{Griffith1993} Griffith M.~R., Wright A.~E., 1993, AJ, 105, 1666
\bibitem[\protect\citeauthoryear{Hotan et al.}{2014}]{Hotan2014} Hotan A.~W., et al., 2014, PASA, 31, e041
\bibitem[\protect\citeauthoryear{Jonker \& Nelemans}{2004}]{Jonker2004} Jonker P.~G., Nelemans G., 2004, MNRAS, 354, 355
\bibitem[\protect\citeauthoryear{Kalemci et al.}{2013}]{Kalemci2013} Kalemci E., Din{\c c}er T., Tomsick J.~A., Buxton M.~M., Bailyn C.~D., Chun Y.~Y., 2013, ApJ, 779, 95
\bibitem[\protect\citeauthoryear{Kennea et al.}{2017a}]{Kennea2017a} Kennea J.~A., Evans P.~A., Beardmore A.~P., Krimm H.~A., Romano P., et al., 2017a, ATel1, 10700
\bibitem[\protect\citeauthoryear{Kennea}{2017b}]{Kennea2017b} Kennea J.~A., 2017b, ATel1, 10731.
\bibitem[\protect\citeauthoryear{Koribalski et al.}{1995}]{Koribalski1995} Koribalski B., Johnston S., Weisberg J.~M., Wilson W., 1995, ApJ, 441, 756
\bibitem[\protect\citeauthoryear{Kreidberg et al.}{2012}]{Kreidberg2012} Kreidberg L., Bailyn C.~D., Farr W.~M., Kalogera V., 2012, ApJ, 757, 36
\bibitem[\protect\citeauthoryear{Kuchar \& Bania}{1990}]{Kuchar1990} Kuchar T.~A., Bania T.~M., 1990, ApJ, 352, 192
\bibitem[\protect\citeauthoryear{Lockman, Blundell, \& Goss}{2007}]{Lockman2007} Lockman F.~J., Blundell K.~M., Goss W.~M., 2007, MNRAS, 381, 881
\bibitem[\protect\citeauthoryear{Mandel}{2016}]{Mandel2016} Mandel I., 2016, MNRAS, 456, 578
\bibitem[\protect\citeauthoryear{Matsuoka et al.}{2009}]{Matsuoka2009} Matsuoka M., et al., 2009, PASJ, 61, 999
\bibitem[\protect\citeauthoryear{McMullin et al.}{2007}]{McMullin2007} McMullin J.~P., Waters B., Schiebel D., Young W., Golap K., 2007, ASPC, 376, 127
\bibitem[\protect\citeauthoryear{Miller et al.}{2018}]{Miller2018} Miller J.~M., et al., 2018, ApJ, 860, L28
\bibitem[\protect\citeauthoryear{Miller-Jones et al.}{2009}]{Miller-Jones2009} Miller-Jones J.~C.~A., Jonker P.~G., Dhawan V., Brisken W., Rupen M.~P., Nelemans G., Gallo E., 2009, ApJ, 706, L230
\bibitem[\protect\citeauthoryear{Murphy et al.}{2007}]{Murphy2007} Murphy T., Mauch T., Green A., Hunstead R.~W., Piestrzynska B., Kels A.~P., Sztajer P., 2007, MNRAS, 382, 382
\bibitem[\protect\citeauthoryear{Negoro et al.}{2017a}]{Negoro2017a} Negoro H., et al., 2017a, ATel1, 10699
\bibitem[\protect\citeauthoryear{Negoro et al.}{2017b}]{Negoro2017b} Negoro H., et al., 2017b, ATel1, 10708
\bibitem[\protect\citeauthoryear{Petrucci et al.}{2008}]{Petrucci2008} Petrucci P.-O., Ferreira J., Henri G., Pelletier G., 2008, MNRAS, 385, L88
\bibitem[\protect\citeauthoryear{Reid et al.}{2011}]{Reid2011} Reid M.~J., McClintock J.~E., Narayan R., Gou L., Remillard R.~A., Orosz J.~A., 2011, ApJ, 742, 83
\bibitem[\protect\citeauthoryear{Reid et al.}{2014}]{Reid2014} Reid M.~J., et al., 2014, ApJ, 783, 130
\bibitem[\protect\citeauthoryear{Russell et al.}{2017a}]{Russell2017a} Russell T.~D., Miller-Jones J.~C.~A., Sivakoff G.~R., Tetarenko A.~J., Jacpot Xrb Collaboration, 2017a, ATel1, 10711
\bibitem[\protect\citeauthoryear{Russell et al.}{2017b}]{Russell2017b} Russell T.~D., Altamirano D., Tetarenko A.~J., Sivakoff G.~R., et al., Jacpot Xrb Collaboration, 2017b, ATel1, 10899
\bibitem[\protect\citeauthoryear{Schlafly \& Finkbeiner}{2011}]{Schlafly2011} Schlafly E.~F., Finkbeiner D.~P., 2011, ApJ, 737, 103
\bibitem[\protect\citeauthoryear{Stiele \& Kong}{2018}]{Stiele2018} Stiele H., Kong A.~K.~H., 2018, ApJ, 868, 71
\bibitem[\protect\citeauthoryear{Tao et al.}{2018}]{Tao2018} Tao L., et al., 2018, MNRAS, 480, 4443
\bibitem[\protect\citeauthoryear{Vahdat Motlagh, Kalemci, \& Maccarone}{2019}]{Vahdat2019} Vahdat Motlagh A., Kalemci E., Maccarone T.~J., 2019, MNRAS, 485, 2744
\bibitem[\protect\citeauthoryear{Wenger et al.}{2018}]{Wenger2018} Wenger T.~V., Balser D.~S., Anderson L.~D., Bania T.~M., 2018, ApJ, 856, 52
\end{thebibliography}
\end{document}